\documentclass[%
 reprint,
 superscriptaddress,
 amsmath,amssymb,
 aps,
floatfix,
]{revtex4-2}

\usepackage{graphicx}% Include figure files
\usepackage{dcolumn}% Align table columns on decimal point
\usepackage{bm}% bold math

\usepackage{verbatim}
\usepackage{physics}
\usepackage{float}
\usepackage{textgreek}
\usepackage{bbold}
\usepackage{lipsum}
\usepackage{placeins}
\usepackage{natbib}
\setcitestyle{square, numbers}
\usepackage[final,hidelinks]{hyperref}

\usepackage{xcolor}
\hypersetup{%
    citebordercolor=green, 
    urlbordercolor=magenta
}
\usepackage[nameinlink]{cleveref}
\crefname{figure}{Fig.}{Figs.}
\crefname{table}{Table}{Tables}
\usepackage{caption}
\newcommand{\mx}[1]{\textup{\textbf{#1}}}

\newcommand{\ctext}[1]{{#1}}

\newcommand{\drv}[1]{\dot{#1}}
\newcommand{\ddrv}[1]{\ddot{#1}}
\newcommand{\tdrv}[1]{\drv{#1}}
\newcommand{\tddrv}[1]{\ddrv{#1}}

\newcommand\sqmatrix[2][c]{%
  \fixTABwidth{T}%
  \setbox0=\hbox{$\tabbedCenterstack{#2}$}%
  \setstackgap{L}{\dimexpr\maxTAB@width+\tabbed@gap}%
  \tabbedCenterstack[#1]{#2}%
}

\newcommand{\RNum}[1]{\uppercase\expandafter{\romannumeral #1\relax}}

\newcolumntype{P}[1]{>{\raggedright\arraybackslash}p{#1\textwidth}}
\newcolumntype{Y}[1]{>{\centering\arraybackslash}p{#1\textwidth}}

\newcounter{Cequ}
\newcounter{Caux}

\usepackage{chngcntr}
\counterwithin*{equation}{section}
\renewcommand{\theequation}{\arabic{section}.\arabic{equation}}

\begin{document}

\preprint{APS/123-QED}

\title{Beyond the Runge-Kutta-Wentzel-Kramers-Brillouin method}% Force line breaks with \\
%\thanks{A footnote to the article %title}%

\author{Jamie Bamber}
\email[]{jarb2@alumni.cam.ac.uk}
\affiliation{Astrophysics Group, Cavendish Laboratory, J.J.Thomson Avenue, Cambridge, CB3 0HE, UK}
\affiliation{Pembroke College, Pembroke Street, Cambridge, CB2 1RF, UK}

\author{Will Handley}
\email[]{wh260@mrao.cam.ac.uk}
\affiliation{Astrophysics Group, Cavendish Laboratory, J.J.Thomson Avenue, Cambridge, CB3 0HE, UK}
\affiliation{Kavli Institute for Cosmology, Madingley Road, Cambridge, CB3 0HA, UK}
\affiliation{Gonville \& Caius College, Trinity Street, Cambridge, CB2 1TA, UK}

%\rule[20pt]{0pt}{0pt}

\date{\rule[10pt]{0pt}{0pt} \today}% It is always \today, today,

\begin{abstract}
We explore higher-dimensional generalizations of the Runge-Kutta-Wentzel-Kramers-Brillouin method for integrating coupled systems of first-order ordinary differential equations with highly oscillatory solutions. Such methods could improve the performance and adaptability of the codes which are used to compute numerical solutions to the Einstein-Boltzmann equations. We test Magnus expansion-based methods on the Einstein-Boltzmann equations for a simple universe model dominated by photons with a small amount of cold dark matter. The Magnus expansion methods achieve an increase in run speed of about $50\%$ compared to a standard Runge-Kutta integration method. A comparison of approximate solutions derived from the Magnus expansion and the Wentzel-Kramers-Brillouin (WKB) method implies the two are distinct mathematical approaches. Simple Magnus expansion solutions show inferior long range accuracy compared to WKB. However we also demonstrate how one can improve on the standard Magnus approach to obtain a new ``Jordan-Magnus'' method. This has a WKB-like performance on simple two-dimensional systems, although its higher-dimensional generalization remains elusive.   
\end{abstract}

\maketitle

%\tableofcontents

\section{\label{sec:Intro}Introduction}

In the new era of high precision cosmology, one of the most important tools in the cosmologist's toolbox is the suite of Boltzmann codes. These methods, including \textsc{class} \cite{CLASS}, \textsc{camb} \cite{CAMB}, and \textsc{pycosmo} \cite{PyCOSMO} provide numerical solutions to the Einstein-Boltzmann equations, which describe the evolution of linear perturbations in the Universe. The dominant computational cost for these codes is integrating coupled systems of linear first order ordinary differential equations (ODEs) with highly oscillatory solutions \cite{CLASS,CAMB}. Currently the codes use several numerical and semianalytic approximations to get around these bottlenecks. However these approximations are designed for specific cosmological theories. A more generalized theory-independent method would make it easier to switch between different models and test new extensions to the current theory. New numerical integration methods could also reduce the run-time of the code. While the time needed to run a single simulation is currently only a few seconds, the many thousands of simulation runs needed for a typical Monte Carlo analysis add up to a substantial computational cost. 

Handley, Lasenby and Hobson proposed in \cite{RKWKB}, and developed with Agocs in \cite{Fruzsina}, a new numerical method based on a combination of the standard Runge-Kutta (RK) approach and the Wentzel-Kramers-Brillouin (WKB) approximation, dubbed the ``Runge-Kutta-Wentzel-Kramers-Brillouin'' (RKWKB) method. In highly oscillatory regions of the solution using the WKB approximation allows integration steps which span many oscillations, giving faster performance than standard RK methods. However RKWKB is limited to essentially one-dimensional systems \cite{RKWKB,Fruzsina}.

We can write a multidimensional coupled system of \mbox{linear} ODEs in the form of \eqref{eqn:Magnus1},
\begin{equation}
    \label{eqn:Magnus1}
    \boldsymbol{\tdrv{x}}(t) = \mx{A}(t) \boldsymbol{x},
\end{equation}
where $\boldsymbol{x}$ is a vector of dynamical variables, $\mx{A}$ is a time-dependent matrix, and an overdot denotes differentiation with respect to time. 

In the mathematics field of numerical analysis there has been substantial work examining the formal solution of \eqref{eqn:Magnus1}, the Magnus expansion \cite{Magnus_appl}. A number of authors have also used the Magnus expansion and related techniques to develop numerical methods to integrate highly oscillatory differential equations \cite{Magnus_appl}, for example Iserles \cite{Iserles2002a,Iserles_2005_OnTheNum}.

These Magnus-based numerical integrators have been extensively applied to problems in quantum and atomic physics \cite{ExplicitMagnus,AdvancesNMR}, as the time dependent Schrodinger equation can be written in the form of (1.1) if we use a basis of wave functions \cite{OnTheExp}. Some alternative methods for solving coupled differential equations have also been explored in the context of primordial cosmology \cite{TwoPoint}. However, Magnus-based methods remain unexplored with respect to the Einstein-Boltzmann equations and cosmological systems in general. Furthermore, to our knowledge, no previous authors have tested these methods with adaptive step size control. 

In this paper we examine the application of Magnus-based techniques to the Einstein-Boltzmann equations. However, we also achieve results which apply to coupled systems of first order ODEs more generally, and hence have relevance to quantum systems, along with problems in many other areas of physics and applied mathematics \cite{RKWKB}.

\section{Background}

\subsection{The WKB approximation}\label{WKB_method}

The WKB method is an example of multiple scale analysis \cite{MultipleScales} applied to oscillatory systems. It is most commonly applied to equations of the form of \eqref{eqn:WKB1} \cite{AMM},
\begin{equation}
    \label{eqn:WKB1}
    \tddrv{x} = -\omega^2(t)x.
\end{equation}
The derivation begins by considering a general solution of the form
\begin{equation}
    \label{eqn:WKB3}
    x = \exp\Big(\sum_{n=0}^{\infty} {S_n}(t)\Big).
\end{equation}
Using the method of separation of scales we can obtain expressions for $S_n$ \cite{AMM}
\begin{align}
    \label{eqn:WKB5}
    S_0 &= \pm i \int_{0}^{t} \omega(t_1) \dd t_1, \\
    \label{eqn:WKB6}
    S_1 &= \textup{const.} - \frac{1}{2}\ln(\omega(t)), \\
    \label{eqn:WKB7}
    2\drv{S}_0 \drv{S}_n + \ddrv{S}_{n-1} + &\sum_{j=1}^{n-1} \drv{S}_j \drv{S}_{n-j}  = 0 \qquad \textup{for } n \geq 2.
\end{align}
Truncating with only the $S_0, S_1$ terms gives the classic WKB approximation:
\begin{equation}
    \label{eqn:WKB8}
    x(t) = \frac{A}{\sqrt{\omega}}e^{i \int_0^t \omega(t_1) \dd t_1} + \frac{B}{\sqrt{\omega}}e^{-i \int_0^t \omega(t_1) \dd t_1}.
\end{equation}
If we include only the $S_0$ term we obtain 
\begin{equation}
    \label{eqn:WKB11}
    \tddrv{x} + \omega^2 x = \pm i \drv{\omega}(t) x,
\end{equation}
implying we require $|\drv{\omega}(t)| \ll |\omega^2(t)|$ for the approximation to be reasonable.

\subsection{The Magnus expansion}\label{Magnus}

The Magnus expansion \cite{Magnus_appl} provides a general solution to equations of the type \eqref{eqn:Magnus1}. As with WKB, it starts by by considering a solution in the form of the exponential of an infinite series \eqref{eqn:Magnus_soln} \cite{Magnus_appl},
\begin{equation}
    \label{eqn:Magnus_soln}
 \boldsymbol{x}(t) = \exp\Big(\sum_{n=1}^{\infty} \mx{\textOmega}_n(t)\Big) \boldsymbol{x}(0),
\end{equation}
the first terms of which are
\begin{align}
    \bm{\Omega}_1(t) &= \int_{0}^{t} \mx{A}(t_1) \dd t_1, \\
    \bm{\Omega}_2(t) &= \frac{1}{2}\int_{0}^{t} \dd t_1 \int_{0}^{t_1} \comm{\mx{A}(t_1)}{\mx{A}(t_2)} \dd t_2, \\
    \begin{split}
    \bm{\Omega}_3(t) &= \frac{1}{6}\int_{0}^{t} \dd t_1 \int_{0}^{t_1} \dd t_2 \int_{0}^{t_2} \Big( \comm{\mx{A}(t_1)}{\comm{\mx{A}(t_2)}{\mx{A}(t_3}} \\
    &+ \comm{\mx{A}(t_3)}{\comm{\mx{A}(t_2)}{\mx{A}(t_1}} \Big) \dd t_3.
    \end{split}
\end{align}
Further terms can be generated by a recursive procedure \cite{Magnus_appl} and involve an increasing number of nested integrals, commutators, and powers of $\mx{A}$.

For the $\bm{\Omega}$ series to converge, $\mx{A}$ must be sufficiently small in some sense. It can be shown \cite{Magnus_appl} that the Magnus series is absolutely convergent for a general complex matrix $\mx{A}$, for $t_0 \leq t \leq T$, if 
\begin{equation}
    \label{eqn:Magnus11}
    \int_{t_0}^T \left \Vert \mx{A}(s) \right \Vert_2 \dd s < \pi,   
\end{equation}
where $\Vert \cdot \Vert_2$ denotes the two-norm of a matrix, which is \cite{two_norm}
\begin{equation}
\label{eqn:Magnus12}
\left\Vert \mx{A}(s) \right\Vert_2 = \textup{max}_{i=1,...,n}\sqrt{\lambda_i(\mx{A}^T\mx{A})},
\end{equation}
where $\lambda_i(\mx{Q})$ denotes the $i$th eigenvalue of matrix $\mx{Q}$. However, the actual convergence domain can be larger than that given by \eqref{eqn:Magnus11} \cite{Magnus_appl}.

\subsection{Stepping numerical integration methods}

A general stepping numerical integration method proceeds by using an approximate solution to the equation to advance from one integration point $\boldsymbol{x}_n$ at time $t_n$ to the next $\boldsymbol{x}_{n+1}$ at time $t_{n+1}$. For a system of first order linear ODEs such as \eqref{eqn:Magnus1} we can write the approximate solution for initial conditions $\boldsymbol{x}(t_0)$ in terms of a solution matrix $\mx{M}(t_0, t)$ \cite{Ward},

\begin{equation}
    \boldsymbol{x}(t) = \mx{M}(t_0, t)\boldsymbol{x}(t_0).
\end{equation}
Then the general stepping procedure is

\begin{equation}
    \label{eqn:gen_step}
    \boldsymbol{x}_{n+1} = \mx{M}(t_n, t_{n+1})\boldsymbol{x}_{n}.
\end{equation}

Runge-Kutta methods work by using truncated Taylor series solutions for $\mx{M}(t_0, t)$. For example, Euler's method, the simplest Runge-Kutta method \cite{RKWKB}, has

\begin{equation}
    \mx{M}(t_0,t) = 1 + (t - t_0)\mx{A}(t_0).
\end{equation}

The RKWKB method (to second order in the WKB series) is obtained by replacing the Taylor series solution with the two independent WKB solutions in \eqref{eqn:WKB8}. As the WKB solutions typically remain good approximations to the true solution over many oscillations, unlike truncated Taylor series solutions, the RKWKB method is able to take much larger steps than Runge-Kutta methods, which in general reduces the required run-time \cite{Fruzsina}. Handley \textit{et al.} \cite{RKWKB} and Agocs \textit{et al.} \cite{Fruzsina} also enable the program to switch to a standard Runge-Kutta-Fehlberg method in slowly oscillating regions; however, in this paper we use RKWKB to refer to the simpler nonswitching approach. 

We can also obtain stepping numerical integration methods of the form of \eqref{eqn:gen_step} from the Magnus expansion. If we can compute the analytic forms of $\mx{\textOmega}_i$ then we can use those directly. In that case the $\mx{\textOmega}$ series truncated at $\mx{\textOmega} = \bm{\Omega}_1 + \cdots + \bm{\Omega}_i$ where $i = 2s-1$ or $2s-2$ for integer $s$ gives a method order $h^{2s+1}$ in the step size $h$ \cite{Magnus_appl}. However, if $\mx{\textOmega}_i$ cannot be computed analytically, we can use an approach from Blanes \textit{et al.} \cite{Magnus_appl} Sec. 5.4. The $\mx{A}$ is Taylor expanded about the midpoint $t_{1/2} = t_n + h/2$ of each step, where $h = t_{n+1} - t_n$ is the step size.
\begin{equation}
    \mx{A}(t) = \sum_{j=0}^{\infty} a_j(t - t_{1/2})^j \quad \textup{where} \quad a_j = \frac{1}{j!} \left. \dv[n]{\mx{A}}{t} \right|_{t=t_{1/2}}.
\end{equation}
One can then use the defining recursion relation for the Magnus expansion to express $\mx{\textOmega}$ to the desired order in $h$. For example to order four we have 
\begin{equation}
    \mx{\textOmega}^{[4]} = \alpha_1 - \frac{1}{12}\comm{\alpha_1}{\alpha_2} + \mathcal{O}(h^5), 
\end{equation}
where $\alpha_j = h^{j}a_{j-1}$. The $\alpha_j$'s can then be approximated using various quadrature rules (including Gauss-Legendre, Simpson, Newton-Cotes) and linear combinations of $\mx{A}$ evaluated at different points in the interval $[t_n, t_{n+1}]$.

\subsection{The Einstein-Boltzmann equations}

In this section we describe the background theory and the Einstein-Boltzmann equations that will be used in Sec. \ref{Num_methods}.

\subsubsection{Overview}

We seek to express the Einstein-Boltzmann equations in the form of \eqref{eqn:Magnus1} \cite{ModernCosmology}. Working in the conformal Newtonian gauge in flat spacetime, the metric is given by 

\begin{equation}
    \dd s^2 = a^2(\eta){[}(1 + 2\Psi)\dd \eta^2 - (1 - 2\Phi)\delta_{ij}\dd x^i \dd x^j{]}.
\end{equation}
where the scalar perturbations are $\Psi, \Phi$, and $\eta$ is conformal time \cite{Ma_Bertschinger}. As we work at early times we neglect dark energy and treat the neutrinos as relativistic. We denote the photon and neutrino temperature perturbations as $\Theta$ and $\mathcal{N}$ respectively. The analogous perturbation $\Theta_P$ describes the strength of the photon polarization. We treat the cold dark matter (CDM) and baryonic matter as nonrelativistic and describe them with overdensities $\delta, \delta_b$ and peculiar velocities $v, v_b$. 

To achieve a linear set of equations we expand the perturbations in spatial Fourier modes, introducing wave number $k$, and we expand the relativistic perturbations in multipole moments $\Theta_l(k, \eta)$ for $l = 0, 1, 2, \ldots, \infty$. The resulting $14 \times 14$ $\mx{A}(k, \eta)$ matrix for $l \leq 2$ is shown in \cref{fig:EBMatrix}.

The conformal Hubble rate is $\mathcal{H} = \tdrv{a}/a$, where an overdot is used to denote the derivative with respect to conformal time. We denote the nonconformal cosmic Hubble rate as $H = \mathcal{H}/a$. The $\tdrv{\tau} = -n_e {\sigma}_T a$ is the conformal time derivative of the optical depth, $\tau$, with $n_e$ the mean electron density and ${\sigma}_T$ the Thompson cross section. Mean densities are described using parameters of the form $\Omega = \rho/\rho_\textup{crit}$ where $\rho$ is the mean density of a component and  $\rho_\textup{crit}$ is the critical density $ = \frac{3H^2}{8\pi G}$. These parameters are denoted $\Omega_{\gamma}, \Omega_{\nu}, \Omega_{\textup{dm}}$ and $\Omega_b$ for the photons, neutrinos, CDM and baryonic matter respectively.

\subsubsection{Photon and CDM universe}

\label{sec:photon_CDM_theory}

As a test case we consider a simplified universe with only photons and cold dark matter. By neglecting the baryonic matter we can avoid the complexities of the Coulomb interaction between the baryons and the photons. We also neglect the photon polarization and all harmonics $l > 1$, approximating the photons as a fluid \cite{ModernCosmology}. In this limit the gravitational potential $\Phi = \Psi$. By making these approximations we obtain a five-dimensional system with $\boldsymbol{x} = [\Theta _0,\Theta _1,\delta,v,\Phi]^{T}$. The reduced dimensions makes this much more amenable than the 14-dimensional system, while still preserving some of the key features of the early universe. By testing on this simplified system we provide a proof-of-principle study for application to the full Einstein-Boltzman equations.  

For adiabatic perturbations and radiation domination we obtain a second order evolution equation for the potential \cite{ModernCosmology,BaumannNotes}
\begin{equation}
    \tddrv{\Phi} + \frac{4}{\eta}\tdrv{\Phi} + \frac{1}{3}{k^2}\Phi = 0.
    \label{eqn:Phi_evo}
\end{equation}
This has a general solution \eqref{eqn:Phi_sol} \cite{ModernCosmology,BaumannNotes} 
\begin{equation}
    \Phi = A_{k}\frac{j_1(x)}{x} + B_{k}\frac{y_1(x)}{x},
     \label{eqn:Phi_sol}
\end{equation}
where $x = k\eta / \sqrt{3}$ and $j_1, y_1$ are first order spherical Bessel functions of the first and second kinds respectively. These take the form 
\begin{equation}
    \begin{split}
    j_1(x) = \frac{\sin(x)}{x^2} - \frac{\cos(x)}{x}, \\
    y_1(x) = -\frac{\cos(x)}{x^2} - \frac{\sin(x)}{x}.
    \end{split}
\end{equation}
As radiation dominates the energy density we also find

\begin{equation}
    \delta_r = 4\Theta_0 = -\frac{2}{3}(k\eta)^2\Phi - 2\eta\tdrv{\Phi} - 2\Phi
    \label{eqn:Theta_Phi},
\end{equation}
where $\delta_r$ is the overdensity for radiation. Hence for $k\eta \gg 1$ we have $\Theta_0 \propto -(k\eta)^2\Phi \sim A_k\cos{(k\eta/\sqrt{3})} + B_k\sin{(k\eta/\sqrt{3})}$. This gives the acoustic oscillations of the photon fluid at the sound speed $c_s=1/\sqrt{3}$ that one would expect. 
The solution in \eqref{eqn:Phi_sol} suggests the potential also oscillates for $k\eta \gtrsim 1$, but with the amplitude decaying as $1/(k\eta)^2$. 

\cleardoublepage
\newpage
\pagebreak

\onecolumngrid

\begin{figure}[H]
\hspace*{-0.7cm}
\centering
\begin{equation*}
\renewcommand\arraystretch{2.0}
\setcounter{MaxMatrixCols}{20}
\begin{bmatrix}
 -2 \mathcal{H} \Omega _{\gamma } & -k & \frac{12 \mathcal{H}^3 \Omega _{\gamma }}{k^2} & 0 & 0 & 0 & -2 \mathcal{H} \Omega _{\nu } & 0 & \frac{12 \mathcal{H}^3 \Omega _{\nu }}{k^2} & -\frac{1}{2}\mathcal{H} \Omega _{\text{dm}} & 0 & -\frac{1}{2} \mathcal{H} \Omega _b & 0 & -\frac{k^2}{3 \mathcal{H}}-\mathcal{H} \\
 \frac{1}{3}k & \tdrv{\tau } & -\frac{4 \Omega _{\gamma } \mathcal{H}^2}{k}-\frac{2}{3}k & 0 & 0 & 0 & 0 & 0 & -\frac{4 \mathcal{H}^2 \Omega _{\nu }}{k} & 0 & 0 & 0 & -\frac{1}{3} i \tdrv{\tau } & \frac{1}{3}k \\
 0 & \frac{2}{5}k & \frac{21}{20}\tdrv{\tau } & \frac{1}{20}\tdrv{\tau } & 0 & \frac{1}{20}\tdrv{\tau } & 0 & 0 & 0 & 0 & 0 & 0 & 0 & 0 \\
 0 & 0 & -\frac{1}{2}\tdrv{\tau } & \frac{1}{2}\tdrv{\tau } & -k & -\frac{1}{2}\tdrv{\tau } & 0 & 0 & 0 & 0 & 0 & 0 & 0 & 0 \\
 0 & 0 & 0 & \frac{1}{3}k & \tdrv{\tau } & -\frac{2}{3}k & 0 & 0 & 0 & 0 & 0 & 0 & 0 & 0 \\
 0 & 0 & -\frac{1}{20}\tdrv{\tau } & -\frac{1}{20}\tdrv{\tau } & \frac{2}{5}k & \frac{19}{20}\tdrv{\tau } & 0 & 0 & 0 & 0 & 0 & 0 & 0 & 0 \\
 -2 \mathcal{H} \Omega _{\gamma } & 0 & \frac{12 \mathcal{H}^3 \Omega _{\gamma }}{k^2} & 0 & 0 & 0 & -2 \mathcal{H} \Omega _{\nu } & -k & \frac{12 \mathcal{H}^3 \Omega _{\nu }}{k^2} & -\frac{1}{2}\mathcal{H} \Omega_{\text{dm}} & 0 & -\frac{1}{2}\mathcal{H}\Omega_b & 0 & -\frac{k^2}{3 \mathcal{H}}-\mathcal{H} \\
 0 & 0 & -\frac{4 \mathcal{H}^2 \Omega _{\gamma }}{k} & 0 & 0 & 0 & \frac{1}{3}k & 0 & -\frac{4 \Omega_{\nu}\mathcal{H}^2}{k}-\frac{2}{3}k & 0 & 0 & 0 & 0 & \frac{1}{3}k \\
 0 & 0 & 0 & 0 & 0 & 0 & 0 & \frac{2}{5}k & 0 & 0 & 0 & 0 & 0 & 0 \\
 -6 \mathcal{H}\Omega_{\gamma} & 0 & \frac{36 \mathcal{H}^3 \Omega _{\gamma }}{k^2} & 0 & 0 & 0 & -6 \mathcal{H} \Omega _{\nu } & 0 & \frac{36 \mathcal{H}^3 \Omega _{\nu }}{k^2} & -\frac{3}{2}\mathcal{H}\Omega_{\text{dm}} & 0 & -\frac{3}{2}\mathcal{H} \Omega_b & 0 & -\frac{k^2}{\mathcal{H}}-3\mathcal{H} \\
 0 & 0 & \frac{12 i \mathcal{H}^2 \Omega _{\gamma }}{k} & 0 & 0 & 0 & 0 & 0 & \frac{12 i \mathcal{H}^2 \Omega _{\nu }}{k} & 0 & -H & 0 & 0 & -i k \\
 -6 \mathcal{H} \Omega _{\gamma } & 0 & \frac{36 \mathcal{H}^3 \Omega _{\gamma }}{k^2} & 0 & 0 & 0 & -6 \mathcal{H} \Omega _{\nu } & 0 & \frac{36 \mathcal{H}^3 \Omega _{\nu }}{k^2} & -\frac{3}{2}\mathcal{H} \Omega_{\text{dm}} & 0 & -\frac{3}{2}\mathcal{H}\Omega_b & -i k & -\frac{k^2}{\mathcal{H}}-3\mathcal{H} \\
 0 & \frac{4 i \tdrv{\tau } \Omega _{\gamma }}{\Omega _b} & \frac{12 i \mathcal{H}^2 \Omega _{\gamma }}{k} & 0 & 0 & 0 & 0 & 0 & \frac{12 i \mathcal{H}^2 \Omega _{\nu }}{k} & 0 & 0 & 0 & \frac{4 \tdrv{\tau } \Omega _{\gamma }}{3 \Omega _b}-H & -i k \\
 -2 \mathcal{H} \Omega _{\gamma } & 0 & \frac{12 \mathcal{H}^3 \Omega _{\gamma }}{k^2} & 0 & 0 & 0 & -2 \mathcal{H} \Omega _{\nu } & 0 & \frac{12 \mathcal{H}^3 \Omega _{\nu }}{k^2} & -\frac{1}{2} \mathcal{H} \Omega _{\text{dm}} & 0 & -\frac{1}{2} \mathcal{H} \Omega _b & 0 & -\frac{k^2}{3 \mathcal{H}}-\mathcal{H} \\
\end{bmatrix}
\end{equation*}
\caption{\ctext{$14 \times 14$ $\mx{A}$ matrix for $l \leq 2$. Here $\boldsymbol{x} = [\Theta _0,\Theta _1,\Theta _2,\Theta _{P0},\Theta _{P1},\Theta _{P2},\mathcal{N}_0,\mathcal{N}_1,\mathcal{N}_2,\delta ,v,\delta_b,v_b,\Phi]^{T}$}}.
\label{fig:EBMatrix}
\end{figure}

\twocolumngrid
To use the numerical methods we must first obtain the relevant $\mx{A}$. Working from the $14 \times 14$ matrix in \cref{fig:EBMatrix} we can derive the reduced $\mx{A}$ matrix shown in \eqref{eqn:pb},
\begin{equation}
\vspace{0.2cm}
\hspace*{0.0cm}
\label{eqn:pb}
\renewcommand\arraystretch{1.5}
\begin{bmatrix}
\tdrv{\Theta}_0 \\
\tdrv{\Theta}_1 \\
\tdrv{\delta} \\
\tdrv{v} \\
\tdrv{\Phi}
\end{bmatrix} = 
\begin{bmatrix}
 -2 \mathcal{H} \Omega _{\gamma } & -k & -\frac{1}{2} \mathcal{H} \Omega _\textup{dm} & 0 & -\frac{k^2}{3 \mathcal{H}}-\mathcal{H} \\
 \frac{k}{3} & 0 & 0 & 0 & \frac{k}{3} \\
 -6 \mathcal{H} \Omega _{\gamma } & 0 & -\frac{3}{2} \mathcal{H} \Omega _\textup{dm} & -i k & -\frac{k^2}{\mathcal{H}}-3\mathcal{H} \\
 0 & 0 & 0 & -H & -i k \\
 -2 \mathcal{H} \Omega _{\gamma } & 0 & -\frac{1}{2} \mathcal{H} \Omega _\textup{dm} & 0 & -\frac{k^2}{3 \mathcal{H}}-\mathcal{H} \\
\end{bmatrix}
\begin{bmatrix}
\Theta_0 \\
{\Theta_1} \\
{\delta} \\
{v} \\
{\Phi}
\end{bmatrix}_{\textstyle.}
\end{equation}
We set the conformal time at matter-radiation equality to be $\eta_{eq} = 1$, set $a(\eta_{eq}) = 1$ and assume $\Omega_{\gamma}(\eta_{eq}) = \Omega_{\textup{dm}}(\eta_{eq}) = 1/2$. This gives
\begin{equation}
\mx{A}(k, \eta) = 
\renewcommand\arraystretch{1.5}
\begin{bmatrix}
 -\frac{1}{\eta } & -k & -\frac{1}{4} & 0 & -\frac{\eta  k^2}{3}-\frac{1}{\eta } \\
 \frac{k}{3} & 0 & 0 & 0 & \frac{k}{3} \\
 -\frac{3}{\eta } & 0 & -\frac{3}{4} & -i k & -\eta  k^2-\frac{3}{\eta } \\
 0 & 0 & 0 & -\frac{1}{\eta^2} & -i k \\
 -\frac{1}{\eta } & 0 & -\frac{1}{4} & 0 & -\frac{\eta  k^2}{3}-\frac{1}{\eta } \\
    \end{bmatrix}_{\textstyle.}
\vspace{2.0cm}
\end{equation}
%

%\cleardoublepage
%\newpage
%\pagebreak

\section{Comparing WKB and the Magnus expansion}

\label{sec:Comparing_Magnus_to_WKB}

\subsection{Comparing analytic solutions}

\label{sec:comparing_analytic}

To investigate the relationship between the WKB approximation and the Magnus expansion we compared the lowest order approximate analytic solutions to \eqref{eqn:WKB1}.

Equation \eqref{eqn:WKB1} can be written in the form of \eqref{eqn:Magnus1} by setting
\begin{equation}
    \label{eqn:Magnus2}
    \boldsymbol{x} = 
    \begin{bmatrix}
        x \\
        \tdrv{x}
    \end{bmatrix}_{\textstyle,}
    \quad \quad 
    \mx{A} = 
    \begin{bmatrix}
        0 & 1 \; \\
        -\omega^2 & 0 \; 
    \end{bmatrix}_{\textstyle.}
\end{equation}

For the Magnus expansion, the first two $\mx{\textOmega}$ terms are given by 
\begin{align}
    \label{eqn:Magnus3}
    \mx{\textOmega}_1 &= 
    \begin{bmatrix}
        0 & t \; \\
        -\int_0^t \omega^2 \dd t_1 \: & 0 \; 
    \end{bmatrix}_{\textstyle,} \\
    \label{eqn:Magnus4}
    \mx{\textOmega}_2 &= 
    \begin{bmatrix}
        \; a & 0 \; \\
        \; 0 & -a \; 
    \end{bmatrix}_{\textstyle,}
\end{align}
where 
\begin{equation}
    a = \tfrac{1}{2} t \int_0^t \omega^2 \dd t_1 - \int_0^t \dd t_1 \int_0^{t_1} \omega^2 \dd t_2 = \tfrac{1}{2}\int_0^t t_1^2 \dv{t} \langle \omega^2 \rangle \dd t_1.
\end{equation}
\\
Including only $\mx{\textOmega}_1$ we obtain $\boldsymbol{x}(t) = \mx{M}(t) \boldsymbol{x}(0)$, where $\mx{M} = \exp(\mx{\textOmega}_1(t))$ given by \eqref{eqn:Magnus5},
\begin{equation}
    \label{eqn:Magnus5}
   \mx{M} = 
   \begin{bmatrix}
        \cos(\sqrt{\langle \omega^2 \rangle}t) & \frac{1}{\sqrt{\langle \omega^2 \rangle}} \sin(\sqrt{\langle \omega^2 \rangle}t) \\
        \sqrt{\langle \omega^2 \rangle} \sin(\sqrt{\langle \omega^2 \rangle}t) & 
        \cos(\sqrt{\langle \omega^2 \rangle}t)
    \end{bmatrix}_{\textstyle.}
\end{equation}
We can then write $x(t)$ as 
\begin{equation}
    \label{eqn:Magnus6}
    x(t) = A\cos(\sqrt{\langle \omega^2 \rangle}t) + B\frac{1}{\sqrt{\langle \omega^2 \rangle}} \sin(\sqrt{\langle \omega^2 \rangle}t).
\end{equation}
The frequency is approximated as $\omega \sim \sqrt{\langle \omega^2 \rangle} = \sqrt{\frac{1}{t} \int_0^t \omega^2 \dd t_1}$. We also get some amplitude correction from the $B$ or $M_{12}$ term. We denote this solution ``first Magnus.'' 
\\
\\
Including both $\mx{\textOmega}_1$ and $\mx{\textOmega}_2$ we obtain
\begin{align}
\label{eqn:Magnus7}
\begin{split}
   \mx{M} &= \mathbb{1} \cos(\hat{\omega}t) + (\mx{\textOmega}_1 + \mx{\textOmega}_2)\frac{1}{\hat{\omega}t}\sin(\hat{\omega}t) \\
   &=
   \begin{bmatrix}
        \; 1 & 0 \; \\
        \; 0 & 1 \;
    \end{bmatrix} \cos(\hat{\omega}t) + 
    \frac{1}{\hat{\omega}t} 
    \begin{bmatrix}
        a & t \; \\
        -\int_0^t \omega^2 \dd t_1 \: & -a \; 
    \end{bmatrix}\sin(\hat{\omega}t)_{\textstyle.}
\end{split}
\end{align}
In \eqref{eqn:Magnus7} the frequency is approximated as $\hat{\omega}$, 
\begin{equation}
    \label{eqn:Magnus8}
    \hat{\omega} = \sqrt{\langle \omega^2 \rangle - (a/t)^2} = \sqrt{\langle \omega^2 \rangle - \frac{1}{4}\langle t^2 \drv{\langle \omega^2 \rangle} \rangle^2},
\end{equation}
where $\drv{\langle \omega^2 \rangle}$ denotes the time derivative of $\langle \omega^2 \rangle$. Then we obtain \eqref{eqn:Magnus9} and \eqref{eqn:Magnus10},
\begin{equation}
    \label{eqn:Magnus9}
    x(t) = A\bigg[\cos(\hat{\omega}t) + \frac{a}{t\hat{\omega}}\sin(\hat{\omega}t)\bigg] + B\frac{1}{\hat{\omega}}\sin(\hat{\omega}t),
\end{equation}
\begin{equation}
    \label{eqn:Magnus10}
    x(t) = A\frac{\sqrt{\langle \omega^2 \rangle}}{\hat{\omega}}\cos(\hat{\omega}t + \hat{\phi}(t)) + B\frac{1}{\hat{\omega}}\sin(\hat{\omega}t),
\end{equation}
with corrections to both the phase and the amplitude, where $\hat{\phi}$ is a time dependent phase term. We denote this solution ``second Magnus.'' For our matrix in \eqref{eqn:Magnus2} $\left\Vert \mx{A}(s) \right\Vert_2 = \textup{max}(\omega^2, 1)$. So \eqref{eqn:Magnus11} implies a convergence domain of $T-t_0 < \pi, \int_{t_0}^T \omega^2 \dd t < \pi$. One can also see from \eqref{eqn:Magnus8} that at sufficiently large times $\hat{\omega}$ becomes imaginary, giving exponentially growing or shrinking solutions. 

For the WKB approximation in Eq. \eqref{eqn:WKB8} the \mbox{$n=0$} term approximates the frequency as 
\begin{equation}
    \label{eqn:WKB9}
    \langle \omega \rangle = \frac{1}{t} \int_{0}^{t} \omega(t_1) \dd t_1, 
\end{equation}
while the \mbox{$n=1$} term provides a correction to the amplitude.
We can write \eqref{eqn:WKB8} in the form $\boldsymbol{x}(t) = \mx{M}(t_0, t)\boldsymbol{x}(t_0)$. Setting $t_0=0$ for simplicity we obtain 
\begin{equation}
    \hspace*{-0.5cm}
    \label{eqn:WKB10}
    \mx{M} = \begin{bmatrix}
    C + \tfrac{\drv{\omega}(0)}{2\omega(0)^2}S & \tfrac{1}{\omega(0)}S\\
    \big(\tfrac{\omega\drv{\omega}(0)}{2\omega(0)^2} - \tfrac{\drv{\omega}}{2\omega}\big)C - \big(\omega + \tfrac{\drv{\omega}\drv{\omega}(0)}{4\omega \omega(0)^2}\big)S & \tfrac{\omega}{\omega(0)}C - \tfrac{\drv{\omega}}{2\omega\omega(0)}S
    \end{bmatrix}_{\textstyle,}
\end{equation}
where $C = \big(\frac{\omega(0)}{\omega}\big)^{\frac{1}{2}}\cos(\langle \omega \rangle t)$ and $S = \big(\frac{\omega(0)}{\omega}\big)^{\frac{1}{2}}\sin(\langle \omega \rangle t)$ are the two independent solutions. Applying the stepping procedure in \eqref{eqn:gen_step} to this $\mx{M}$ gives you the RKWKB method (to second order) \cite{RKWKB}.

We can thus observe that the Magnus expansion and the WKB method take fundamentally different approaches for solving the same problem. The classic WKB approximation approximates the oscillation frequency as $\langle \omega \rangle$, while the first Magnus solution uses the root mean square, $\sqrt{\langle \omega^2 \rangle}$. The terms of the WKB series involve successively higher order derivatives of $\omega$, while the successive Magnus expansion terms involve successively higher order integrals. 

The WKB approximation is specialized to highly oscillatory solutions, and fails when $\omega$ goes to zero. The Magnus expansion can cope with nonoscillatory solutions and is less specialized, which comes at the cost of inferior long range accuracy compared to WKB, as we shall see in Sec. \ref{sec:Example_eq}.

\subsection{\label{sec:Example_eq}Testing on trial equations}

These solutions were plotted for two trial equations taken from Handley \textit{et al.} \cite{RKWKB}, namely the Airy equation and the ``burst'' equation, which both have  analytic exact solutions. The Airy equation takes the form
\begin{equation}
    \label{eqn:Airy_eq}
    \tddrv{x}(t) = -tx,
\end{equation}
and the burst equation is given by
\begin{equation}
    \label{eqn:burst_eq}
    \tddrv{x} = -\omega^2(t)x \quad , \quad \omega^2(t) = \frac{n^2-1}{(1+t^2)^2},
\end{equation}
where $n$ is an integer.

\begin{figure}
    \hspace*{-0.5cm} 
    \centering
    \includegraphics[trim={0 0.4cm 0 0},clip]{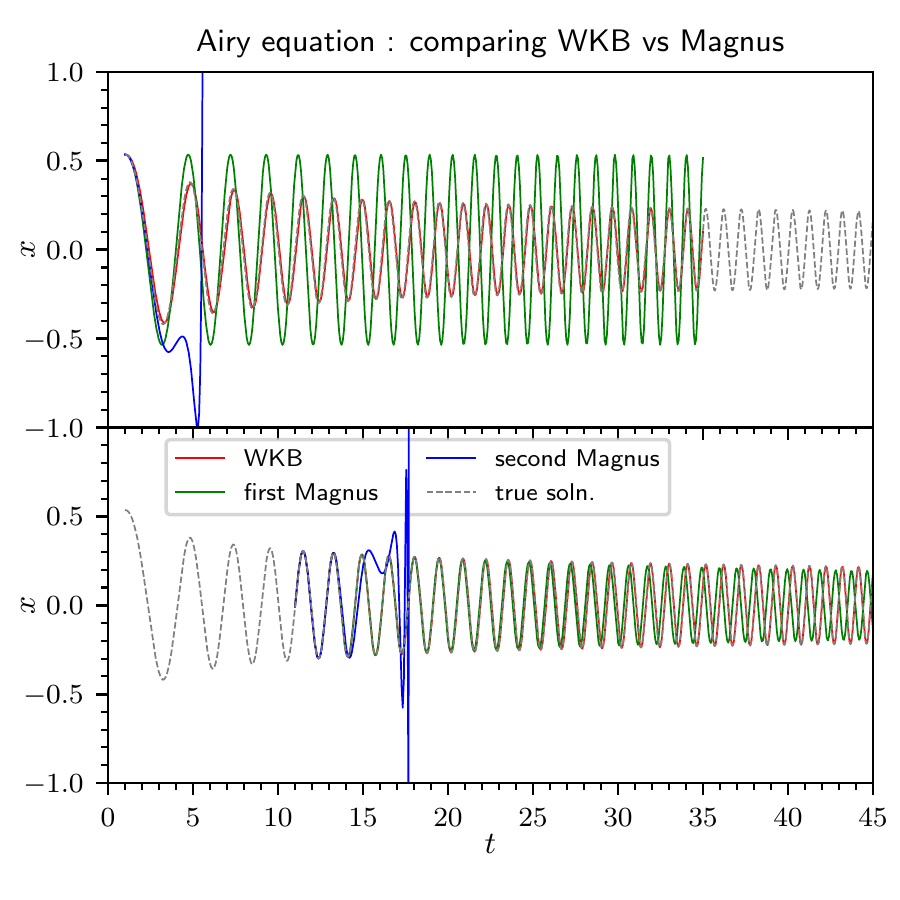}
    \caption{\ctext{WKB, and the $\mx{\textOmega}_1$ and $\mx{\textOmega}_1 + \mx{\textOmega}_2$ Magnus analytic solutions to the Airy equation. TOP: from $t_0 = 1$ to $t = 35$. BOTTOM: from $t_0 = 11$ to $t = 45$.}} 
    \label{fig:Airy_M1_M2_WKB}
\end{figure}

The analytic WKB approximate solution \eqref{eqn:WKB8} and the first and second Magnus solutions \eqref{eqn:Magnus6} and \eqref{eqn:Magnus9} for the Airy equation \eqref{eqn:Airy_eq} are plotted in \cref{fig:Airy_M1_M2_WKB}.
For the upper plot, with $t_0 = 1$, the WKB solution remains a very good approximation to the true solution; however, the first and second Magnus solutions perform poorly. The first Magnus reproduces the increase in oscillation frequency, but does not reproduce the changes in amplitude. The second Magnus solution breaks down completely around $t-t_0 \sim \pi$ showing the exponential growth predicted in Sec. \ref{sec:comparing_analytic}.

%\begin{figure}[h]
%    \hspace*{-0.5cm} 
%    \centering
%    \includegraphics[width=0.55\textwidth%]{Analytic_test_Airy_WKB_M1_M2__t_start=1%0_t_stop=45index=0.pdf}
%    \caption{\ctext{WKB, and the %$\mx{\textOmega}_1$ and %$\mx{\textOmega}_1 + \mx{\textOmega}_2$ %Magnus solutions to the Airy equation %with $t_0 = 10$}} 
%    \label{fig:Airy_shifted_M1_M2_WKB}
%\end{figure}

If we shift the starting point to $t=11$, as in the bottom plot in \cref{fig:Airy_M1_M2_WKB}, the first Magnus solution follows the true solution more closely. However unlike the WKB solution it fails to follow the frequency of the true oscillation  at large times. The second Magnus solution still breaks down as predicted.

\begin{figure}
    \hspace*{-1.0cm} 
    \centering
    \includegraphics[trim={0 0.0cm 0 0},clip]{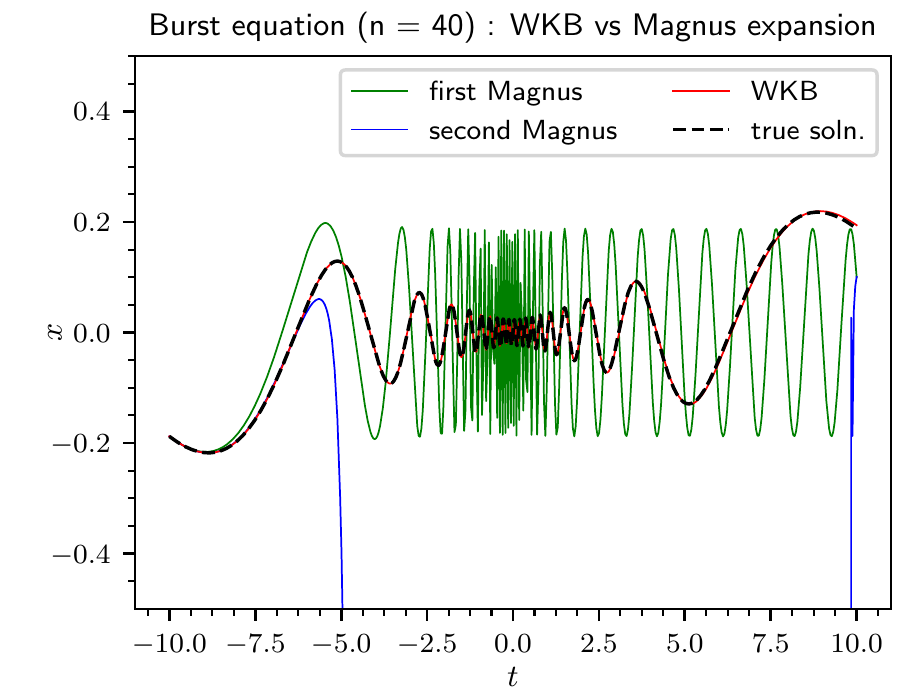}
    \caption{\ctext{WKB, and the $\mx{\textOmega}_1$ and $\mx{\textOmega}_1 + \mx{\textOmega}_2$ Magnus solutions to the burst equation, $n=40$, analytic solutions plotted for $t=-10$ to $t=+10$}} 
    \label{fig:burst_M1_M2_WKB}
\end{figure}

Next we consider the burst equation \eqref{eqn:burst_eq}. Figure \ref{fig:burst_M1_M2_WKB} shows the analytic WKB, first Magnus and second Magnus approximate solutions from $t_0=-10$ up to $t=+10$ with $n=40$. While WKB provides an excellent match to the true solution throughout the whole range, both Magnus solutions give very poor approximations. The first Magnus solution fails to follow the changes in amplitude or frequency, and the second Magnus solution once again diverges after a relatively short period.
 
\begin{figure}
    \hspace*{-1.0cm} 
    \centering
    \includegraphics[trim={0 0.0cm 0 0},clip]{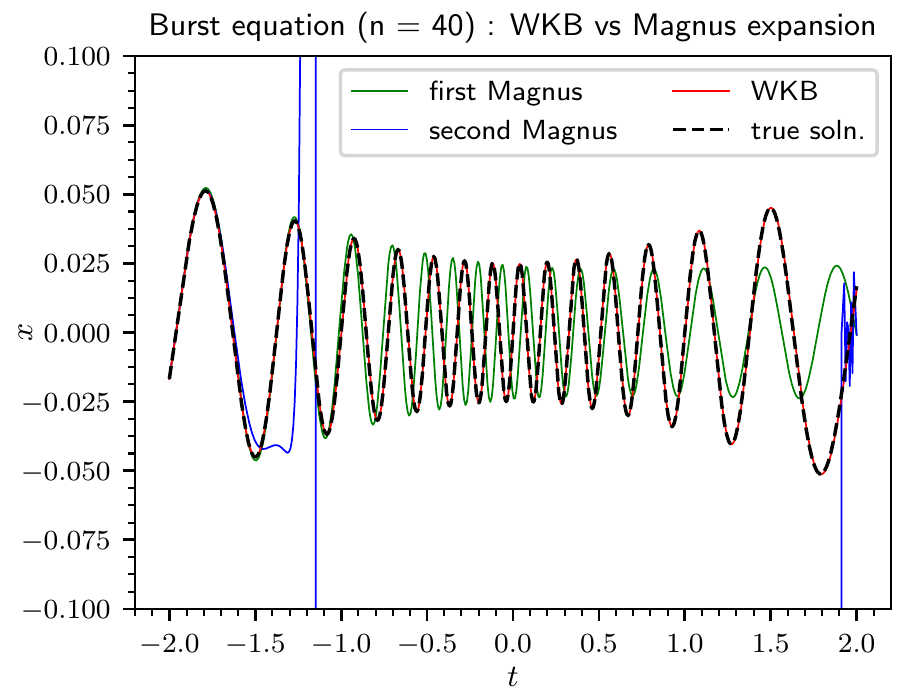}
    \caption{\ctext{WKB, and the $\mx{\textOmega}_1$ and $\mx{\textOmega}_1 + \mx{\textOmega}_2$ Magnus solutions to the burst equation, $n=40$.  Solutions plotted for the central region, $t=-2$ to $t=+2$ (note the change in scale). The solutions were again computed analytically.}} 
    \label{fig:burst_M1_M2_WKB_centre}
\end{figure}

Figure \ref{fig:burst_M1_M2_WKB_centre} shows the result if we try to apply the WKB and Magnus solutions to only the small, central, highly oscillatory region. This region has a smaller $\drv{\omega}/\omega$ ratio, so it should be easier to model. Even then the WKB solution dramatically outperforms both Magnus solutions. 

\subsubsection{A closer look at the Magnus error}\label{Magnus_closer_error}

To show why the first Magnus solution compares so poorly in terms of accuracy we can differentiate the Magnus solutions. Let $C_M$ and $S_M$ be the two independent first Magnus solutions,
\begin{align}
\begin{split}
    C_M &= \cos(\sqrt{\langle \omega^2 \rangle}t), \\
    S_M &= \frac{1}{\sqrt{\langle \omega^2 \rangle}} \sin(\sqrt{\langle \omega^2 \rangle}t).
\end{split}
\end{align}
Let $1 + \epsilon=\omega^2/\langle \omega^2 \rangle$ and $\xi = \drv{\omega}/\omega$, and then
\begin{equation}
\begin{split}
    \tddrv{C}_M = &-\omega^2\left(1 + \frac{\epsilon^2}{4(1 + \epsilon)}\right)C_M \;\dots \\
    &\dots - \omega^2\xi S_M - \frac{\omega^2}{t}\left( \epsilon - \frac{3\epsilon^2}{4(1+\epsilon)} \right) S_M,
\end{split}
\end{equation}
\begin{equation}
\begin{split}
    \tddrv{S}_M = &-\omega^2\left(1 + \frac{\epsilon^2}{4(1 + \epsilon)}\right)S_M \;\dots \\
    &\dots -\bigg[ \xi\epsilon - \frac{1}{4t}(2\epsilon + \epsilon^2) \bigg] \left( \frac{1}{t} S_M - C_M \right).
\end{split}
\end{equation}
One can see that for $\epsilon = 0, \xi = 0$ we get back \eqref{eqn:WKB1}. As the errors can increase with $\epsilon$ as well as $\xi$, they are potentially much larger than for WKB.

\section{The Jordan-Magnus method}
\label{sec:JM}

The WKB approximation provides good long range accuracy but is limited to one-dimensional systems. In contrast the truncated Magnus expansion provides a matrix-based approach suitable for multidimensional systems, but with inferior long range accuracy as we saw in Sec. \ref{sec:Comparing_Magnus_to_WKB}. Ideally we want a way to combine the positive aspects of both WKB and Magnus, that is achieve WKB-like long range accuracy with a solution that works on multidimensional systems. In this section we introduce a new approach based on the Magnus expansion, dubbed the ``Jordan-Magnus'' method, which goes some way toward this goal.  

If the $\mx{A}$ matrix were time-independent, we could solve the system by diagonalizing $\mx{A}$ to obtain the independent oscillating modes. Consider introducing a linear transformation to \eqref{eqn:Magnus1}. Let $\boldsymbol{x}_P = \mx{P}^{-1}\boldsymbol{x}$, and let $\mx{J} = \mx{P}^{-1}\mx{A}\mx{P}$ for some transformation matrix $\mx{P}(t)$. Then

\begin{equation}
    \label{eqn:AP}
    \tdrv{\boldsymbol{x}}_P = \mx{A}_P\boldsymbol{x}_P = \big[\mx{J} + \tdrv{(\mx{P}^{-1})}\mx{P}\big]\boldsymbol{x}_P,
\end{equation}
where $\tdrv{(\mx{P}^{-1})}$ denotes the time derivative of $\mx{P}^{-1}$. We want to choose a $\mx{P}$ such that $\mx{A}_P$ is as diagonal as possible. Let $\mx{A}_P = \bm{\Lambda} + \mx{K}$ where $\bm{\Lambda}$ is diagonal and $\mx{K}$ has zeros on the diagonal. 

Once again we seek a solution in the form $\boldsymbol{x}(t) = \mx{M}(t)\boldsymbol{x}(0)$.
We can obtain an approximate solution by neglecting $\mx{K}$, 
\begin{equation}
    \label{eqn:lambda_only}
    \mx{M}(t) \approx \mx{P}(t)\exp\bigg(\int^t_{0} \mx{\textLambda} \dd t_1\bigg)\mx{P}^{-1}(0).
\end{equation}
This avoids the potentially costly matrix exponential. Alternatively we can include $\mx{K}$ and obtain
\begin{equation}
    \label{eqn:LK}
    \mx{M}(t) \approx \mx{P}(t)\exp\bigg(\int^t_{0} (\mx{\textLambda} + \mx{K}) \dd t_1\bigg)\mx{P}^{-1}(0),
\end{equation}
which is the first Magnus solution to \eqref{eqn:AP}. 

An obvious choice of linear transformation is one that diagonalizes $\mx{A}$. Now in general $\mx{A}$ may not have a diagonal form; however, all $N \times N$ matrices can be placed in Jordan normal form \cite{JordanForm} (which reduces to diagonal form for diagonalizable matrices). 

For $\mx{A} = \big[\begin{smallmatrix}
    0   &   1 \; \\
    -\omega^2   &   0 \;
    \end{smallmatrix}\big]$, $\mx{A}$ is indeed diagonalizable, and we obtain 
\begin{align}
    \mx{J} &= \begin{bmatrix}
     -i\omega & 0 \\
     0  & i\omega  
    \end{bmatrix}_{\textstyle,} & \;\; 
    \mx{P} = \begin{bmatrix}
     i/\omega & -i/\omega \\
     1  & 1  
    \end{bmatrix}_{\textstyle,} \\
    \mx{\textLambda} &= \begin{bmatrix}
     \frac{\drv{\omega}}{2\omega}-i\omega & 0 \\
     0  & \frac{\drv{\omega}}{2\omega} + i\omega  
    \end{bmatrix}_{\textstyle,} &\;\; 
    \mx{K} = \begin{bmatrix}
     0 & -\frac{\drv{\omega}}{2\omega} \\
     -\frac{\drv{\omega}}{2\omega}  & 0  
    \end{bmatrix}_{\textstyle.}
\end{align}
We then find the $\bm{\Lambda}$ only solution \eqref{eqn:lambda_only} returns
\begin{align}
    \label{eqn:JM1sol}
    \begin{split}
    \mx{M}(t) &= \begin{bmatrix}
     \Big(\frac{\omega(0)}{\omega(t)}\Big)^{\frac{1}{2}}\cos(\langle \omega \rangle) & 
     \frac{1}{\omega(0)}\Big(\frac{\omega(0)}{\omega(t)}\Big)^{\frac{1}{2}}\sin(\langle \omega \rangle) \\
     - \omega(0)\Big(\frac{\omega(0)}{\omega(t)}\Big)^{\frac{1}{2}}\sin(\langle \omega \rangle) & \Big(\frac{\omega(t)}{\omega(0)}\Big)^{\frac{1}{2}}\cos(\langle \omega \rangle)
    \end{bmatrix} \\
    &= 
    \begin{bmatrix}
     C & 
     \frac{1}{\omega(0)}S \; \\
     - \omega(0)S & \frac{\omega}{\omega(0)}C \;
    \end{bmatrix}_{\textstyle,}
    \end{split}
\end{align}
where $C$ and $S$ are the independent WKB solutions from Sec. \ref{WKB_method}. Thus we have obtained a WKB-like framework from a Magnus-like approach. 

The result is not identical to the WKB solution matrix \eqref{eqn:WKB10}. We have no $\drv{\omega}$ terms, and indeed because the Magnus expansion proceeds by integration as opposed to differentiation, extending the series does not introduce any further derivatives. This means that, while similar, the Jordan-Magnus approach remains distinct from WKB. In certain senses this lack of $\drv{\omega}$ terms may be an advantage as derivatives can be difficult to evaluate numerically \cite{Fruzsina}.

\begin{figure}
    \hspace*{-0.5cm} 
    \centering
    \includegraphics[trim={0 0.0cm 0 0},clip]{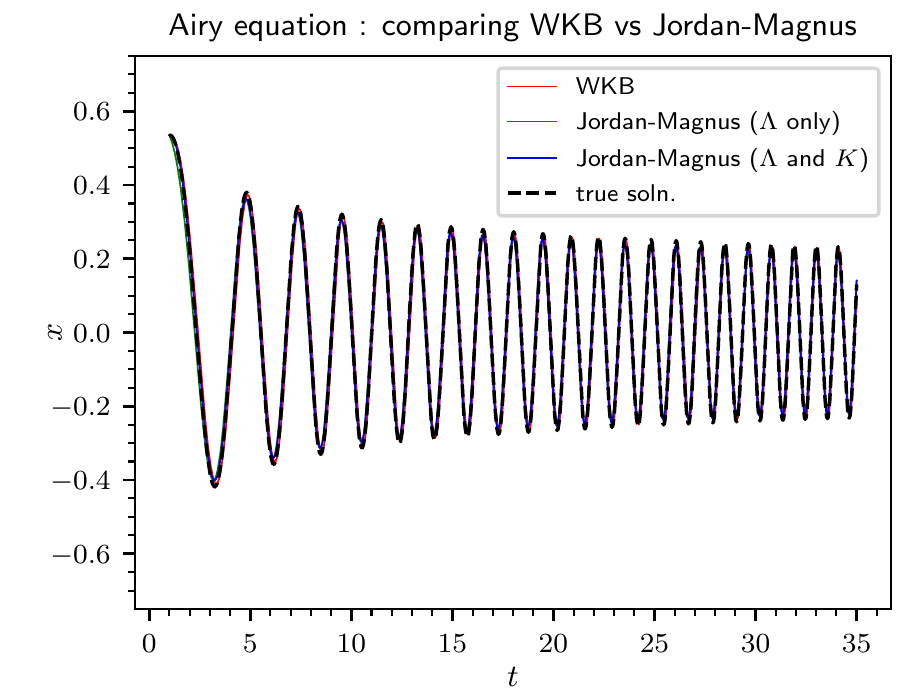}
    \caption{\ctext{WKB, and the $\bm{\Lambda}$ only and $\bm{\Lambda} + \mx{K}$ Jordan-Magnus analytic solutions to the Airy equation with $t_0=1$. Note all three lines lie very close on top of one another and the true solution.}}
    \label{fig:Airy_WKB_JM1l_JM1lk}
\end{figure}

\begin{figure}
    \hspace*{-0.5cm} 
    \centering
    \includegraphics[trim={0 0.0cm 0 0},clip]{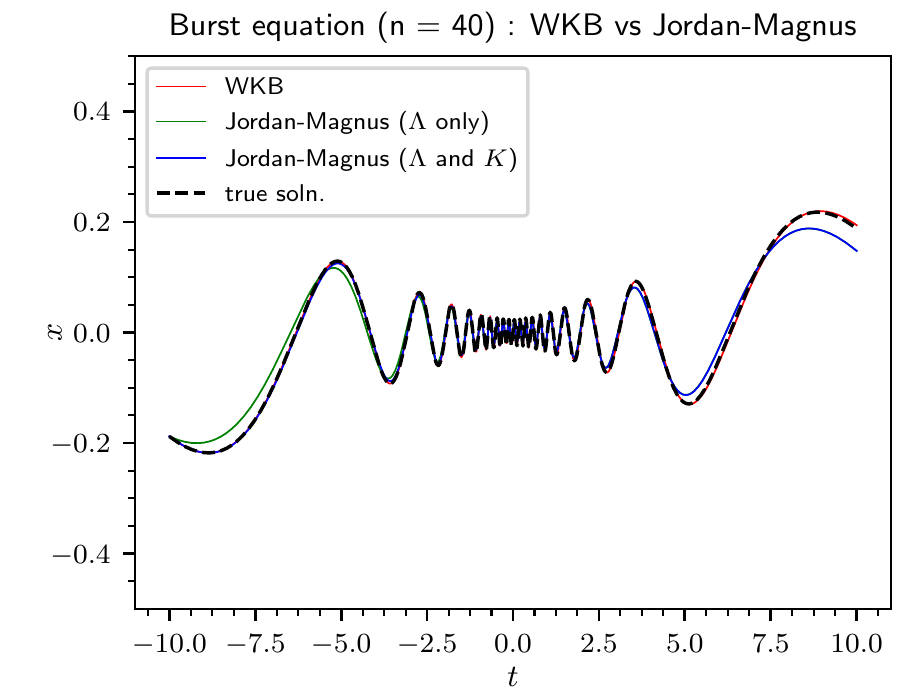}
    \caption{\ctext{WKB, and the $\bm{\Lambda}$ only and $\bm{\Lambda} + \mx{K}$ Jordan-Magnus solutions to the burst equation with $n=40$, analytic solutions plotted for $t=-10$ to $t=+10$. Note the Jordan-Magnus lines lie on top of one another for most the graph.}} 
    \label{fig:burst_WKB_JM1l_JM1lk}
\end{figure}

If we try plotting the results of the Jordan-Magnus method for the Airy and burst equations (\cref{fig:Airy_WKB_JM1l_JM1lk,fig:burst_WKB_JM1l_JM1lk}) we can see it dramatically outperforms the first and second Magnus solutions in accuracy, achieving results similar to the WKB solution. 

However, this method also has some disadvantages over standard Magnus methods. It is ultimately reliant on being able to obtain $\mx{P}, \bm{\Lambda}$, and $\mx{K}$ as functions of $t$. In addition, numerical Jordan decomposition is inherently unstable \cite{JordanForm}. This implies we need to compute the analytic, symbolic, Jordan normal form of $\mx{A}$ to apply the method. While this is possible for the two-dimensional \eqref{eqn:Magnus1} system, it quickly becomes prohibitive for more complicated systems with more dimensions.

While there is no stable numerical algorithm for obtaining Jordan-normal form, algorithms do exist for diagonalization \cite{Matrix_Comp}. Furthermore, although a complex square matrix is not guaranteed to be diagonalizable, a very large proportion of such matrices are diagonalizable \cite{AEMatrix}. Hence diagonalization algorithms could form the basis for a more fully numerical Jordan-Magnus method, although at present efficient methods remain elusive.

\section{Results}

\label{Num_methods}

A \textsc{python} code was developed \cite{MyCode} to implement linear numerical integration methods of the form of \eqref{eqn:gen_step} using the WKB, truncated Magnus, and Jordan-Magnus approximate solutions. The \textsc{numpy} package \cite{numpy} was used to perform the numerical calculations, and the \textsc{sympy} package \cite{sympy} was used to perform the symbolic manipulation. For comparison a RKF4(5) method and a second order (in the WKB series) RKWKB method were also implemented. The method of adaptive step size control used is described in Appendix \ref{adapt_step}.

Figure \ref{fig:Airy_burst_num} shows the result of the numerical code for the different methods on the Airy and burst equations described in Sec. \ref{sec:Example_eq}. For both equations we can see that the RKF4(5) method requires the smallest step size, and the Magnus methods use roughly similar intermediate step sizes. As expected the RKWKB method gives the largest step sizes, quickly reaching the maximum allowed step size $h_{\textup{max}}$, although it also gives the largest relative error. The Jordan-Magnus method gives smaller step sizes than RKWKB, but larger step sizes compared to the other methods.

Each plot includes a note of $T$, the time taken for the integration for each method. This of course varies substantially depending on the machine and the code implementation. However the ratio between the run-times for the different methods should give a reasonable indication of the relative computational cost. \cref{tab:ratios} shows the averaged results for \cref{fig:Airy_burst_num}. We can see that the methods with the largest step size give the smallest time ratio and thus the smallest computational cost. The RKWKB method performs best and Jordan-Magnus second best. This result also supports the approach taken in the previous sections of seeking an approximate solution with good long range accuracy, as this allows large integration step sizes. 

We note that the relative errors shown in \cref{fig:Airy_burst_num} are not representative of what could be achieved by an optimized code (for example see Agocs \textit{et al.} \cite{Fruzsina} for a practical implementation of RKWKB to fourth order in the WKB series). Nonetheless, the results concerning the relative step sizes and relative run-times achieved should still apply.

\

\begin{table}[H]
\centering
\renewcommand\arraystretch{1.5}
\begin{tabular}{l| l l}
Method & Airy eq. & Burst eq. ($n=40$)\\
\hline
\hline
First Magnus & $\; 0.115 \pm 0.004 \;\;$ & $0.78 \pm 0.02$ \\
Magnus 4$^{\circ}$ GL & $\; 0.244 \pm 0.004 \;\;$ & $0.81 \pm 0.01$ \\
Second Magnus & $\; 0.253 \pm 0.003 \;\;$ & $0.503 \pm 0.008$ \\
RKWKB $\;$ & $\; 0.0029 \pm 0.0003 \;\;$ & $0.0065 \pm 0.0004$ \\
Jordan-Magnus $\;$ & $\; 0.050 \pm 0.007 \;\;$ & $0.246 \pm 0.028$
\end{tabular}
\caption{\ctext{Ratio of integration times vs. RKF4(5) time for the different plots. Averaged over 20 runs.}}
\label{tab:ratios}
\end{table}

\onecolumngrid

\begin{figure}[H]
    \hspace*{-0.5cm} 
    \centering
    \begin{tabular}{@{}c@{}}
    \includegraphics{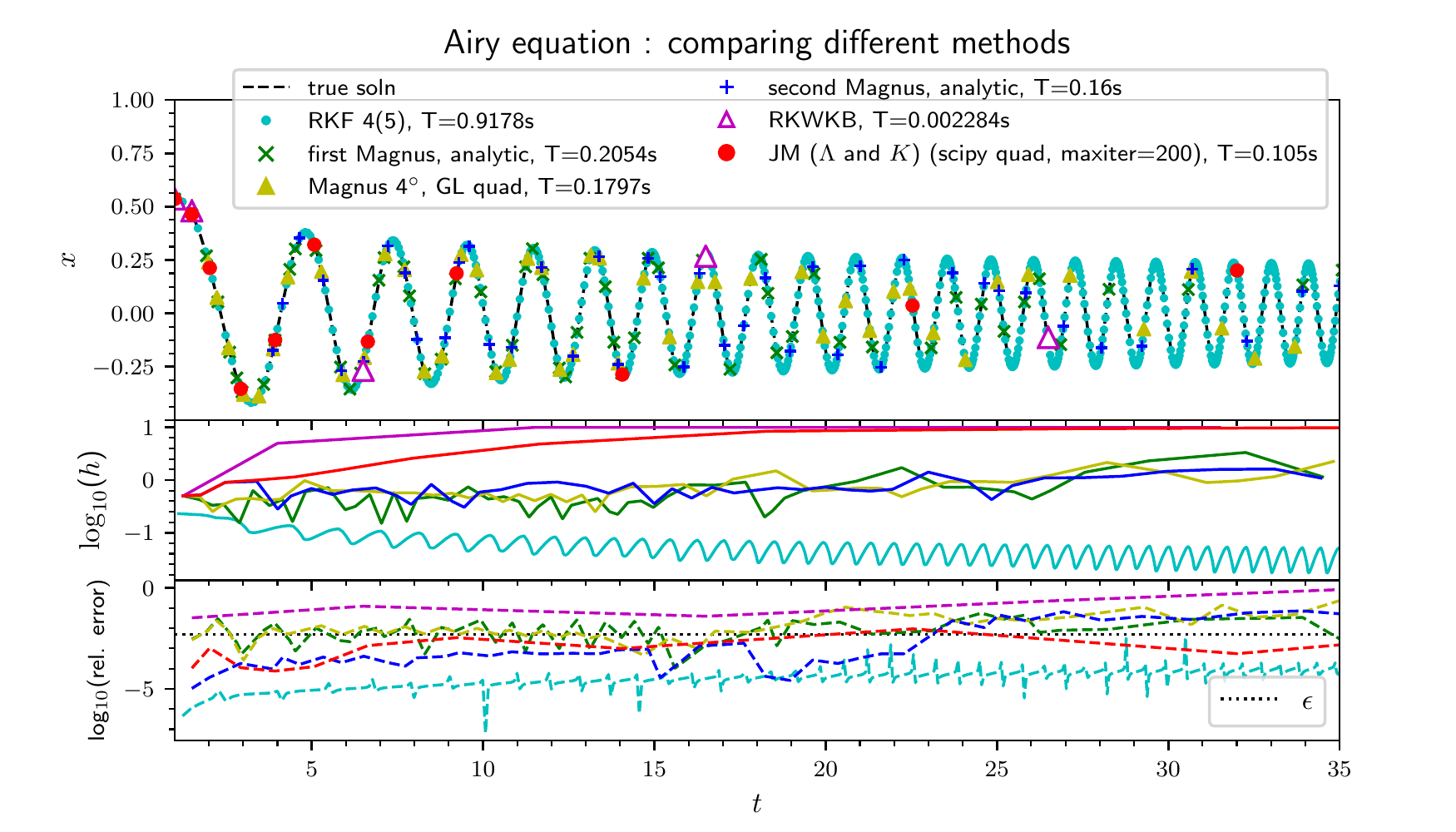}
    \end{tabular}
    
    \vspace*{-0.5cm}
    
    \begin{tabular}{@{}c@{}}
    \includegraphics{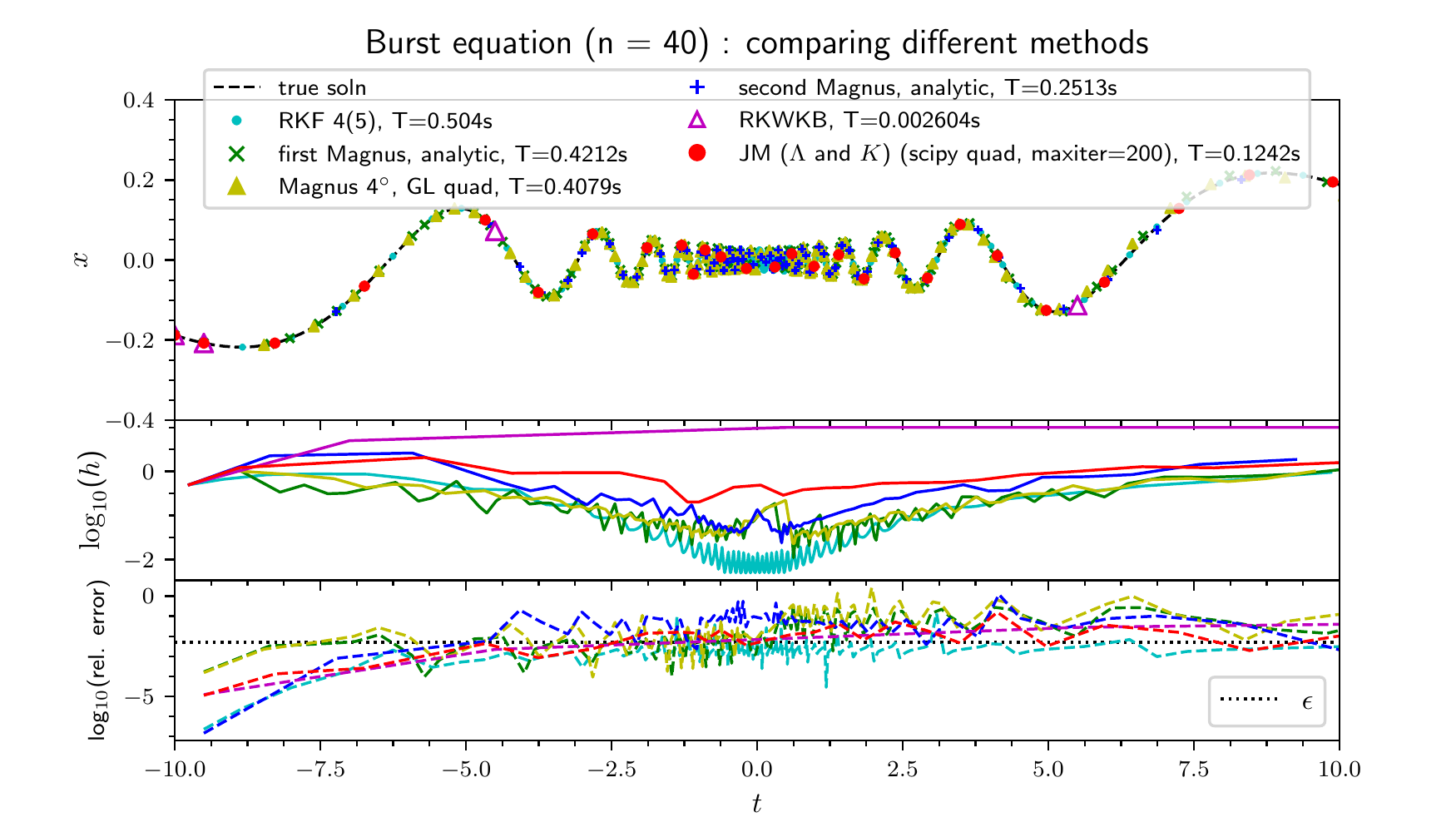}
    \end{tabular}
    \caption{\ctext{Applying numerical integration with adaptive step size control to the Airy equation and burst equation with $n=40$. For the Magnus and RKWKB solutions $\epsilon = 0.005$, $a_{\textup{tol}} = 0.005$, $r_{\textup{tol}} = 1$. For the RKF solution, $\epsilon_{RKF} = 0.005$, $a_{\textup{tol}} = 1$, $r_{\textup{tol}} = 2$. Initial step size $h_0 = 0.5$, $h_{\textup{max}} = 10$, $h_{\textup{min}} = 0.01$ for Magnus methods. For RKF4(5) $h_{\textup{max}} = 2.5$, $h_{\textup{min}} = 0.005$. In each plot the layout is TOP: the integration results. MIDDLE: $\log_{10}|h|$ where $h$=step size. BOTTOM: $\log_{10}|\textup{relative error}|$. The $\epsilon$ line shows the target error of $\epsilon = 0.005$.}}
    
    \label{fig:Airy_burst_num}
    
\end{figure}

\cleardoublepage
\newpage
\pagebreak

\twocolumngrid

\FloatBarrier

%\cleardoublepage
%\newpage
%\pagebreak

%\cleardoublepage
%\newpage
%\pagebreak

\onecolumngrid

\begin{figure}[ht]
    \vspace*{+0.0cm}
    \hspace*{-0.5cm} 
    \centering
    \includegraphics{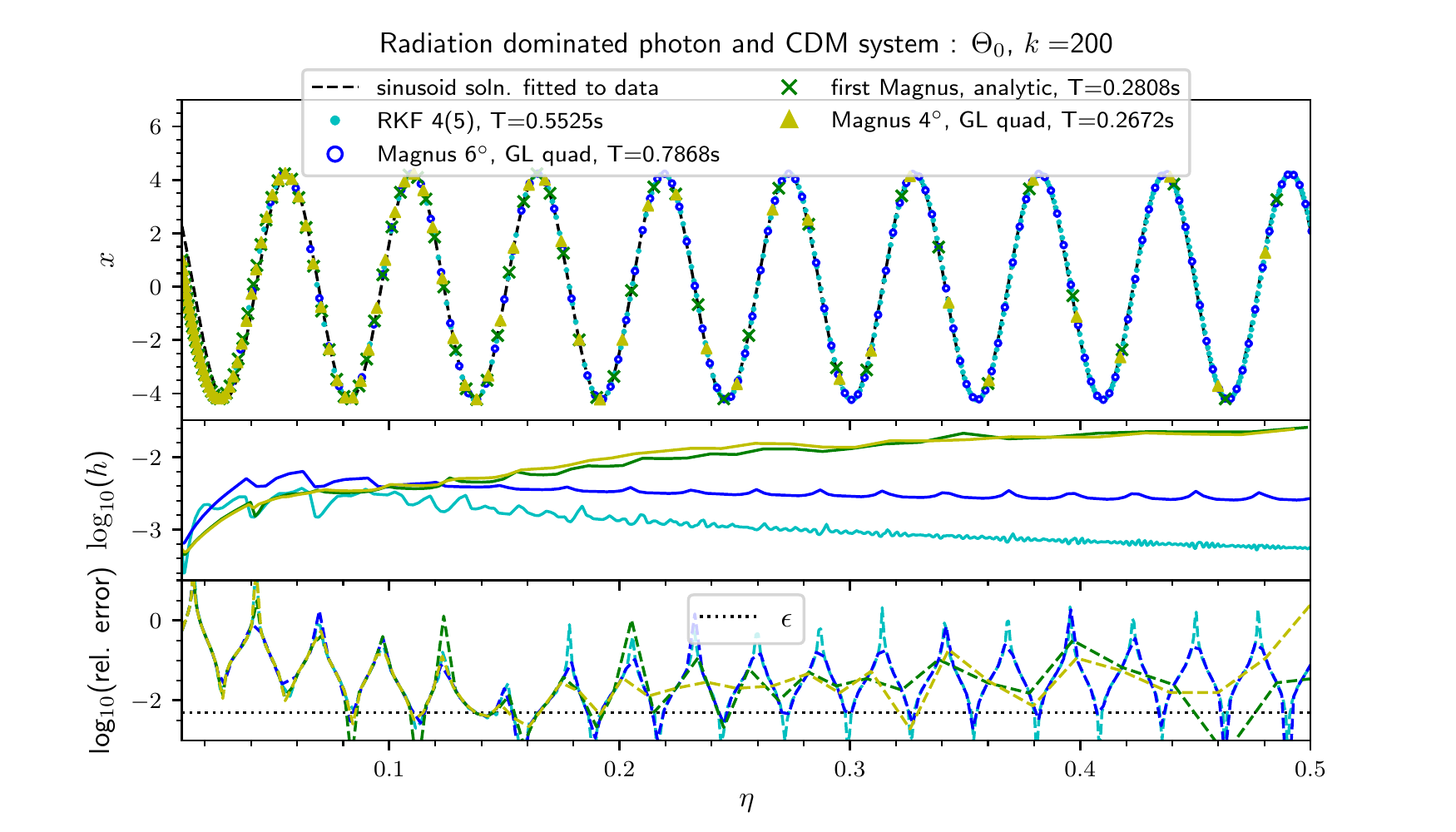}
    \caption{\ctext{Applying numerical integration methods to the photon \& CDM system. The black dashed line is the analytic solution for large $k\eta$ which is $C_0\cos{(k(\eta + C_1)}/\sqrt{3})$ where $C_0, C_1$ are coefficients fitted to the RKF45 data. The initial conditions are $\eta_0 = 0.01$, $\boldsymbol{x}_0 = [1, 2, 1, 2, 1]$. The parameters are $k=200$, $h_0=0.025$, $h_\textup{min} = 0.00025$, $h_{\textup{max}} = 2.5$, $\epsilon = 0.005$. For the Magnus methods $a_{\textup{tol}} = 0.005$, $r_{\textup{tol}} = 1$, for the RKF method $a_{\textup{tol}} = 4$, $r_{\textup{tol}} = 2$. The layout is as in \cref{fig:Airy_burst_num}.}}
    \label{fig:Theta0_num}
\end{figure}

\twocolumngrid

\subsection{Application to the Einstein-Boltzmann equations: Photon and CDM universe}
\label{sec:PhotonOsc}

Finally we applied the methods and numerical integration code to the Einstein-Boltzmann equations for a simple radiation dominated universe with only photons and CDM, as described in Sec. \ref{sec:photon_CDM_theory}. 

\begin{figure}[H]
    \hspace*{-0.5cm} 
    \centering
    \includegraphics{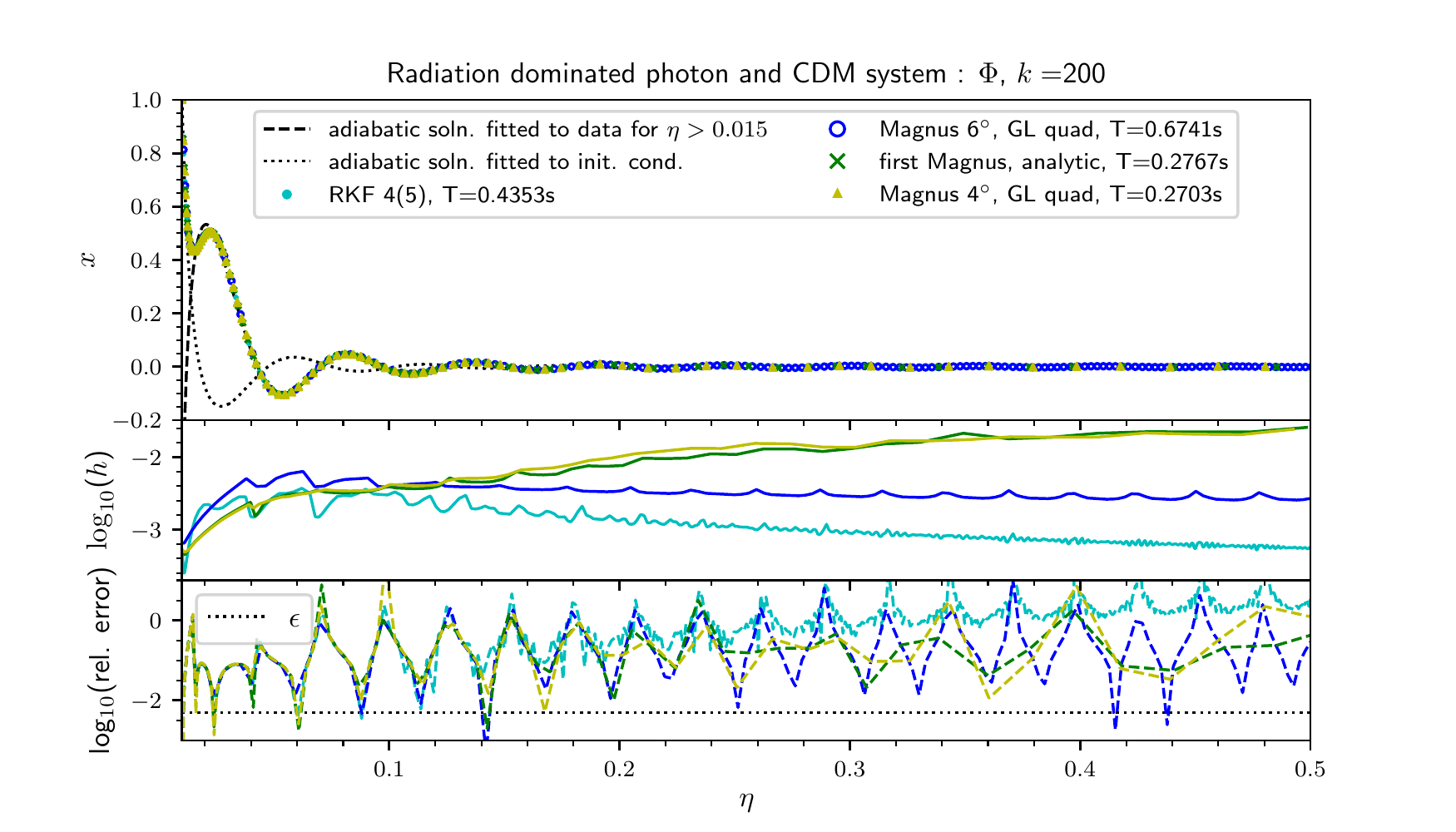}
    \caption{\ctext{Applying numerical integration methods to the photon and CDM system, showing potential $\Phi$. The black dotted line is the adiabatic solution \eqref{eqn:Phi_sol} fitted to the initial conditions, and the black dashed line is the adiabatic solution fitted to the RKF data for $\eta \geq 0.05$. The initial conditions and all other parameters are as in \cref{fig:Theta0_num}}} 
    \label{fig:Phi_num}
\end{figure}

Unfortunately it was not possible to use the new Jordan-Magnus method on this system due to the difficulty of obtaining a symbolic, analytic Jordan normal form. However we can test the analytic first order Magnus solution, fourth and sixth order numerical Magnus solutions, and the RKF4(5) method. The results for $\Theta_0$ and $\Phi$ are shown in \cref{fig:Theta0_num,fig:Phi_num}. 

Fitting a solution of the form $A_0\cos{(k\eta}/\sqrt{3}) + B_0\sin{(k\eta}/\sqrt{3})$, where $A_0$ and $B_0$ are constants, to the results gives good agreement for \cref{fig:Theta0_num} for all but the earliest times as expected. 

%\cleardoublepage
%\newpage
%\pagebreak

For \cref{fig:Phi_num}, fitting the adiabatic solution for $\Phi$ \eqref{eqn:Phi_sol} to the initial conditions gives good agreement only for very early times up to about $\eta \approx 0.015$. However, a separate adiabatic solution \eqref{eqn:Phi_sol} fitted to the RKF4(5) data for $\eta > 0.015$ gives good agreement for $\eta \gtrsim 0.03$ onwards. Hence while the results do not fit a single adiabatic solution they can be well described by two adiabatic regimes with a transition region between $0.015 \lesssim \eta \lesssim 0.03$.

The plots of $\log_{10}(h)$ show RKF4(5) gives the smallest step size, the sixth  order GL Magnus method has intermediate step sizes, and the first Magnus and fourth order GL Magnus methods give the largest step sizes.

As in Sec. \ref{Num_methods} we can compare the average ratio of run-times (\cref{tab:ratios_PhotonOsc}). This shows that the sixth order GL Magnus method actually underperforms RKF4(5), with the smaller number of steps not compensating for the larger time per step. However, the analytic first Magnus and fourth order GL Magnus methods both give improved performance compared to RKF4(5). 

\begin{table}[h]
\centering
\renewcommand\arraystretch{1.5}
\begin{tabular}{P{0.2}| Y{0.2}}
Method & Ratio \\
\hline
\hline
Magnus 6$^{\circ}$ GL & $\; 1.58 \pm 0.06 \;\;$ \\
First Magnus & $\; 0.71 \pm 0.03 \;\;$ \\
Magnus 4$^{\circ}$ GL & $\; 0.62 \pm 0.03 \;\;$
\vspace*{-4cm}
\end{tabular}
\caption{\ctext{Ratio of integration times for different methods vs. RKF4(5) for the photon \& CDM system. Parameter settings and initial conditions are as in \cref{fig:Theta0_num,fig:Phi_num}. Averaged over 20 runs.}}
\label{tab:ratios_PhotonOsc}
\end{table}

This shows that despite their inferior performance compared to WKB in Sec. \ref{sec:Comparing_Magnus_to_WKB}, Magnus expansion based methods can still outperform RKF4(5) on a real cosmological higher-dimensional system.

\onecolumngrid

\twocolumngrid

\section{Conclusion}
\label{sec:disc}

\label{sec:conc}

In this work we sought to explore methods for efficiently integrating coupled systems of first order ODEs with highly oscillatory solutions, which would be a higher-dimensional extension of the RKWKB method, with particular applications to cosmological systems of interest.

The work of previous authors \cite{Iserles2002a,Iserles_2005_OnTheNum,RKWKB,Fruzsina} suggests methods based on the Magnus expansion could be candidates for a multidimensional equivalent of RKWKB. However comparison of the analytic Magnus expansion and WKB approximate solutions implies the two methods are fundamentally distinct mathematical approaches. For equations of the form of \eqref{eqn:WKB1} WKB approximates the frequency as $\langle \omega \rangle$, whereas the first Magnus solution approximates it as $\sqrt{\langle \omega^2 \rangle}$. The WKB solution series involves successively higher derivatives of $\omega$ while the Magnus expansion involves successively higher order integrals. Testing on trial equations of the form of \eqref{eqn:WKB1}, the WKB solution give much higher accuracy. This indicates that the Magnus expansion and WKB based methods are orthogonal lines of research; however, the relationship between the two may benefit from more formal mathematical inquiry.

The simplest Magnus expansion based approximate solutions were found to gave inferior long range accuracy compared to WKB. By introducing a linear transformation we can convert the $\mx{A}$ matrix to a more diagonal form, obtaining a new ``Jordan-Magnus'' method. This gave excellent long range accuracy on simple two-dimensional problems which closely resembled the WKB solutions. However, currently the method relies on finding the symbolic Jordan normal form of $\mx{A}$ which is very difficult for high-dimensional cases. It could form the basis for a more fully numerical method which could be applied to arbitrarily high dimensions. 

A \textsc{python} code was created to implement the corresponding numerical integration methods, incorporating an adaptive step size procedure and the RKWKB and RKF4(5) methods for comparison. The code leaves plenty of room for optimization, and one could certainly improve the performance by switching to a compiled language like \textsc{c}. On simple two-dimensional systems, simple Magnus expansion based numerical methods ran significantly faster than RKF4(5), with the Jordan-Magnus method running faster still, although not achieving the speed of RKWKB.

Finally we applied this code and these methods to a simplified set of Einstein-Boltzmann equations corresponding to photon and CDM oscillations in the radiation era. The results again demonstrate that Magnus expansion based methods performed better than RKF4(5) by giving a speed improvement of about $50\%$. However, the improvement was not as impressive as for the simpler systems, and we could not implement the Jordan-Magnus method. For large times the numerical results fit the expected solutions for adiabatic perturbations. However, we could not fit a single adiabatic solution to the numerical results for the potential $\Phi$. Instead the results suggest a transition between two different adiabatic regimes. This could indicate a breakdown of the adiabatic approximation for that system.

Future work would involve testing these Magnus expansion based methods on other multidimensional systems, including the $14 \times 14$ Einstein-Boltzmann matrix for $l \leq 2$. If successful, one could extend up to high $l$ values and see how one might include these methods in future Boltzmann solver codes. 

These results demonstrate that multidimensional numerical integration methods based on the Magnus expansion show promise for improving the performance of not only cosmological Boltzmann solvers but also many similar problems in other areas of physics.

\section*{Acknowledgements}

J.B. thanks the University of Cambridge Physics Part III project scheme, which formed the basis for this work. W.H. was supported by a
Gonville \& Caius research fellowship. We are also grateful for the advice of Fruzsina Agocs and Lukas Hergt of the Kavli Institute for Cosmology, Cambridge.

\bibliography{BRKWKB}% Produces the bibliography via BibTeX.

\onecolumngrid

\appendix

\cleardoublepage
\newpage
\pagebreak

\section{Procedure for adaptive step size control}

\label{adapt_step}

We implemented the adaptive step size procedure laid out in Press and Flannery (1988) \cite{NR}. To estimate the error, and thus the correct step size for each step, we compute the estimate of $\boldsymbol{x}_{n+1}$ from one step of size $h$, and a separate estimate using two steps size $h/2$, denoting this $\boldsymbol{x}^{*}_{n+1}$. The error is then $\Delta_i = |x_{i \; {n+1}}^{*} - x_{i \; n+1}|$ for each component $i$.  We define a maximum tolerable error as $\Delta^{\textup{max}}_i = \epsilon(a_{\textup{tol}} + r_{\textup{tol}}|x_{i\;n+1}^*|)$, where $a_{\textup{tol}}$ and $r_{\textup{tol}}$ parametrize the absolute and relative error tolerance, and $\epsilon$ is a small scale factor.

The procedure is then as follows:
\begin{enumerate}
    \item Calculate $\boldsymbol{x}^{*}_{n+1}$ and $\boldsymbol{x}_{n+1}$ for step size $h$.
    
    \item Calculate $\Delta_i$ and $\Delta^{\textup{max}}_i$ and the error ratio $R = \sqrt{\frac{1}{N}\sum_i^N (\Delta_i/\Delta^{\textup{max}}_i)^2}$.
    
    \item If $R \leq 1$, compute a new $h$ as $h_{\textup{new}} = hSR^{-1/(\nu+1)}$ where $S$ = safety factor $\sim 0.99$, $\nu$ = order of method. 
    
    Go to 5.
    
    \item Else if $R > 1$, compute a new $h$ as $h_{\textup{new}} = hSR^{-1/\nu}$.
    
    If $h_{\textup{new}} < h_{\textup{min}}$, set $h_{\textup{new}} = h_{\textup{min}}$ and go to 5. 
    
    Otherwise go to 1.
    
    \item If $h_{\textup{new}} < 0.2h$, set $h_{\textup{new}} = 0.2h$.
    
    If $h_{\textup{new}} > h_{\textup{max}}$, set $h_{\textup{new}} = h_{\textup{max}}$.
    
    If $h_{\textup{new}} > 10h$, set $h_{\textup{new}} = 10h$.
    
    \item Use current $\boldsymbol{x}^{*}_{n+1}$ as the estimate. Advance time. Set $h = h_{\textup{new}}$. Go to next step.
\end{enumerate}    

The RKWKB method used the adaptive step size procedure outlined above, setting $\nu = 1$. The RKF4(5) was implemented following \cite{NR}. The approach to step size control is the same, except the error is calculated using the difference between the fourth and fifth order RK methods.

The $a_{\textup{tol}}$ and $r_{\textup{tol}}$ values for the RKF4(5) and Magnus methods were chosen using trial and error to achieve the largest step sizes for the target error $\epsilon$ when applied to the burst equation as in \cref{fig:Airy_burst_num}. For the RKWKB method the values for the Magnus methods were used.

\end{document}